\documentclass[1p]{elsarticle}
\usepackage{lipsum}
\makeatletter
\def\ps@pprintTitle{%
 \let\@oddhead\@empty
 \let\@evenhead\@empty
 \def\@oddfoot{}%
 \let\@evenfoot\@oddfoot}
\makeatother
\usepackage{lineno,hyperref}

\usepackage{graphicx} 
\usepackage{amsmath} 
\usepackage{amssymb}  
\usepackage{amsthm}
\usepackage{dsfont}
\usepackage{float}
\usepackage{color}
\usepackage{transparent}
\usepackage{multirow}
\usepackage{natbib}
\usepackage{subcaption}
\usepackage{tikz}
\usepackage{booktabs}

\usetikzlibrary{shapes,arrows}
\usepackage{verbatim}

\modulolinenumbers[5]

\newtheorem{prop}{Proposition}

\newtheorem{definition}{Definition}

\newproof{pf}{Proof}

\newcommand{\x}[0]{{\bf x}}

\begin{document}

\begin{frontmatter}

\title{Exponential stability of a PI plus reset integrator controller by a sampled-data system approach}

\author[zero]{M. A. Dav{\'o}\corref{mycorrespondingauthor}}
\cortext[mycorrespondingauthor]{Corresponding author}
\ead{Miguel.Davo-Navarro@gipsa-lab.fr}

\author[First,Second]{F. Gouaisbaut}
\ead{fgouaisb@laas.fr }

\author[Third]{A. Ba{\~n}os}
\ead{abanos@um.es}

\author[First,Fourth]{S. Tarbouriech}
\ead{sophie.tarbouriech@laas.fr }

\author[First,Fourth]{A. Seuret}
\ead{aseuret@laas.fr }

\address[zero]{Gipsa-lab, Grenoble Campus, 11 rue des math{\'e}matiques, BP 46, 38402 Saint Martin d'H{\'e}res Cedex, France.}
\address[First]{CNRS, LAAS, 7 avenue du Colonel Roche, F-31400 Toulouse, France.}
\address[Second]{Univ de Toulouse, UPS, LAAS, F-31400, Toulouse, France.}
\address[Third]{Dpto. Inform\'atica y Sistemas, University of Murcia, 30071 Murcia, Spain.}
\address[Fourth]{Univ de Toulouse, LAAS, F-31400, Toulouse, France.}
\tnotetext[]{\copyright {\ }2016. This manuscript version is made available under the CC-BY-NC-ND 4.0 license http://creativecommons.org/licenses/by-nc-nd/4.0/}

\begin{abstract}
The paper deals with the stability analysis of time-delay reset control systems, for which the resetting law is assumed to satisfy a time-dependent condition. A stability analysis of the closed-loop system is performed based on an appropriate sampled-data system. New linear matrix inequality (LMI) conditions are proposed to ensure the exponential stability of the closed-loop resulting of the connection of a plant with a proportional and integral controller together with a reset integrator (PI+RI).
\end{abstract}


\end{frontmatter}


\section{Introduction}

\textcolor{black}{Reset control systems represent a particular class of hybrid dynamical systems \cite{TeelBook12},
in which the system state (or part of it) is reset at the
instants it intersects some reset surface. When the reset 
is defined as a function of time, one can consider the reset
control system as an impulsive system, in which some states of the controller are subjected to impulsive actions (they are usually set to zero) at some instants.} The research field of reset control systems is \textcolor{black}{rather} focused on feedback control with the main aim of overcoming the performance of linear control systems while keeping low the order of the system and its complexity. Since the seminal work of J. C. Clegg in 1958 (\cite{r_C1958}), a multitude of works have brought to light the benefits of the reset compensation over linear feedback control schemes (e.g., \textcolor{black}{\cite{r_KH1974,r_HR1975,r_BHCH2001,r_NZT2008,r_FNZ2011,r_BB2012-book,r_GWXLG2011,r_ZNTW2014,Fichera:ACC12,FicheraIJRNC12,r_HDTP2013}}). Despite the significant progress on reset control systems, only few works have considered reset control strategies for processes in presence of time-delay (\cite{r_FBCB2011,r_BV2012,r_MGNBB2013,r_BPC2014,r_BPTZ2014}). In particular, it has been shown that a proportional and integral (PI) controller may be improved without increasing the cost of feedback (sensitivity to sensor noise) by simply adding in parallel a Clegg integrator (CI) \cite{r_KH1974} resulting in a reset controller known as PI+CI \cite{r_BV2012}. The benefits of this controller highlight specially in integrating processes \cite{r_MDB2015}. \textcolor{black}{Although the PI+CI was originally designed to perform reset actions when the error signal is zero, several modification of the resetting law have been propounded in the literature with the aim of improving the performance of the closed-loop system (see, e.g., \cite{r_BB2012-book}): fixed reset band, variable reset band, etc. In order to encompass the most general resetting law, this article concerns with a generalization of the PI+CI controller in which the Clegg integrator is replaced with a general reset integrator. This controller will be referred to as PI+RI (RI stands for reset integrator), and its reset actions can be specified by an arbitrary time/state-dependent resetting law, instead of only the zero crossing resetting law (reset actions are performed when the input is zero).}

Although, several real-world applications (see for instance \cite{r_VB2010,r_DB2013-2,r_DDBB2014,r_BD2014}) have shown the practical advantages of the PI+CI controller (particular case of the PI+RI for a kind of resetting laws), there does not exist results studying the stability of time-delay reset control systems comprising this controller. Recently, different results on the stability of time-delay reset control systems have been developed: delay-independent criteria \cite{r_BB2009,r_BPTZ2014}, delay-dependent criteria \cite{r_BB2010,r_PBDT2012,r_DB2013}, uncertain reset surface \cite{r_GX2012}, input-to-output stability \cite{r_MCB2013,r_MDB2013}. In general, these stability results are based on the Lyapunov-Krasovskii (LK) theorem and functionals, and involve two conditions; the first one (flow condition) imposes constraints over the derivative of a functional between the reset instants and the other condition (jump condition) deals with the instantaneous changes of the functional due to the reset actions. The main limitation of these results comes from the flow condition, since it requires the stability of the base system (system without reset actions). The challenge in the stability analysis of the PI+RI controller arises from its structure of two integrators in parallel, that leads into a non-minimal realization of its base system, and makes more difficult to fulfill the flow condition. 

This work focuses on providing exponential stability criteria of a time-delay reset control system whose feedback controller is a PI+RI. The main idea of this work is the formulation of the reset integrator into the framework of sampled-data systems (\cite{F2014}), which allows representing the time-delay reset control system as a sampled-data system. In this way, exponential stability of the reset control system can be concluded from the analysis of the resulting sampled-data system. The advantages of this approach are twofold: a minimal realization of the PI+RI controller is obtained, and the solutions of the sampled-data system are continuous which leads into a continuous LK functional, i.e., no jump condition is required. 
\textcolor{black}{The stability analysis of sampled-data systems has been performed in many works using approaches related to robust analysis \cite{fujioka2009stability}, a time-delay modelling \cite{fridman2004robust}, an impulsive system interpretation \cite{fridman2010refined,naghshtabrizi2008exponential}, the use of looped-functionals \cite{S2012} or a clock-dependent Lyapunov approach \cite{briat2015theoretical}.} However, these methods cannot be directly applied into the sampled-data system under study. This is due to its particular structure in which the plant is also a closed-loop system with an internal time-delay. The stability criterion provided in this work is based on the extension of the results of \cite{S2011,S2012}, and its combination with a new LK functional provided in \cite{SG2014}. \textcolor{black}{Furthermore, this work can be viewed
as a comprehensive version of the conference work \cite{davo/adhs2015}, where only the asymptotic stability of the closed-loop system has been ensured. The paper can be also complementary to \cite{liuteel} due to the type of LK functionals to be considered and the temporal nature of the proposed conditions.}

This article is organized as follows. Section \ref{sec:problem_statement} formulates the problem and establishes an equivalence between reset control and sampled-data systems. Section \ref{sec:exponential_stability} presents several propositions on exponential stability of time-delay reset control systems by means of a sampled-data system formulation. Some examples of applications of the proposed criterion are given in Section \ref{sec:Examples}. Finally, Section \ref{sec:conclusions} concludes the article.

{\em Notation}: Throughout the article, the sets $\mathds{N}$, $\mathds{R}$, $\mathds{R}^n$, $\mathds{R}^{n\times n}$ and $\mathbb{S}^n$  denote the sets of natural numbers, real numbers, the n-dimensional Euclidean space and the set of $n\times n$ matrices and symmetric matrices of dimension $n\times n$, respectively. A column vector is denoted by ${\bf x} \in \mathds{R}^{n}$ and ${\bf x}^\top$, its transpose. Given two vectors ${\bf  x}_1$ and ${\bf x}_2$, we write $({\bf x}_1,{\bf x}_2)$ to denote $[{ \bf x}_1^\top { \bf x}_2^\top]^\top$.  The notation $\| \x\|$ is the Euclidean norm for $\x\in \mathds{R}^n$. $\mathcal{C}([a,b],\mathds{R}^n)$ stands for the set of continuous functions mapping $[a,b]$ to $\mathds{R}^n$. The identity matrix and the zero matrix of adequate dimensions are denoted by $I$ and $0$, respectively. The notation $P>0$ for $P\in \mathbb{S}^{n}$ means that $P$ is positive definite ($P<0$ means negative definite). The set of symmetric positive definite matrices is denoted by $\mathds{S}_+^n$. For a matrix $A\in \mathds{R}^{n\times n}$, the notation $He(A)$ refers to $A+A^\top$. For any matrices $A,B$, the notation $diag(A,B)$ stands for the block diagonal matrix $\left[\begin{smallmatrix} A&0\\0&B\end{smallmatrix}\right]$.

\section{Problem statement}
\label{sec:problem_statement}

The system under study is the standard feedback control system, which is formed by a plant P and a PI+RI reset controller. The plant is considered to be a linear and time invariant (LTI) system without direct feedthrough from the input to the output. The minimal state-space description of the plant is given by
\begin{equation}\label{eq:Planta}
P :\left\lbrace \begin{array}{l}
\dot{\x}_{p}(t) = A_{p} \x_{p}(t) + B_{p} u_p(t),\\
y_p(t) = C_{p} {\bf x}_{p}(t),
\end{array}	
\right.
\end{equation}
\noindent where $\x_p \in \mathds{R}^{n_p}$ and $A_p \in \mathds{R}^{n_p \times n_p}$, $B_p \in \mathds{R}^{n_p \times 1}$, $C_p \in \mathds{R}^{1 \times n_p}$.

\textcolor{black}{Originally, the Clegg integrator is a nonlinear integrator whose state is reset to zero when its input is zero \cite{r_C1958}. However, the term Clegg integrator has been also used to refer to nonlinear integrators with more general state-dependent resetting laws. In this work, we consider the term {\em reset integrator} (RI) to refer to a nonlinear integrator which is reset to zero at those instants specified by an arbitrary time/state-dependent resetting law. The parallel connection of a PI controller and a reset integrator (RI) leads into the PI+RI controller (similarly to the PI+CI controller). In addition, it is assumed that for a particular resetting law there exists for every execution of the system an strictly increasing sequence of nonzero instants defined by $$\mathds{T}=\{t_k\}_{k\in\mathbb N},$$ 
for which the following assumptions stand}
\begin{equation}\label{def_T}
t_0=0,\qquad T_k:= t_{k+1}-t_k\in [\mathcal T_m \ \mathcal T_M],\qquad \lim_{k\rightarrow +\infty}t_k=+\infty,
\end{equation}
where $0\leq \mathcal T_m\leq \mathcal T_M$ are positive numbers. The last equation means that $\mathds{T}$ has no accumulation points and that Zeno phenomenon is avoided.
A state-space realization of the PI+RI is obtained by using a state $\mathbf{x}_r = ( x_i, x_{ri}) \in \mathds{R}^2$, where $ x_i$ is the integral term state and  $x_{ri}$ corresponds to the reset integrator state. As a result, the PI+RI is described by the following impulsive differential equation
\begin{equation}
\label{eq:PICI}
PI+RI:\left\{ \begin{array}{ll}
 \dot{\mathbf{x}}_r(t) = B_r u_r(t),  & (t,\x_r(t),u_r(t)) \notin \mathcal{R},  \\ 
 \mathbf{x}_r(t^+)=A_\rho \mathbf{x}_r(t), & (t,\x_r(t),u_r(t)) \in \mathcal{R}, \\
y_r(t) = C_r\mathbf{x}_r(t) + k_p u_r(t),
\end{array}	
\right.															
\end{equation}
for some resetting set $\mathcal{R}$ and where $B_r=k_i \left[\begin{smallmatrix} 1\\1\end{smallmatrix}\right]$, $C_r=\left[ \ 1-p_r \ \ p_r \ \right]$, $A_\rho = \left[\begin{smallmatrix} 1&0\\0&0\end{smallmatrix}\right]$, and $$\x_r(t^+)=\lim_{\begin{smallmatrix} \epsilon \rightarrow 0\\ \epsilon>0\end{smallmatrix}} \x_r(t+\epsilon)$$ is referred to as {\em after-reset state}. Besides the tuning parameters of the PI controller ($k_p$ and $k_i$), the PI+RI possesses an extra dimensionless parameter $p_r \in [0,1]$, referred to as the {\em reset ratio} (as the PI+CI controller). This parameter represents the part of the whole integral term over which the reset action is applied. 

Note that if the resetting set $\mathcal{R}$ does not guaranteed the assumption on $\mathds{T}$, then a time regularization technique (see, e.g., \cite{JELS1999}) can be applied by augmenting the PI+RI with an extra state. 

The feedback interconnection of the plant P and the PI+RI controller is considered to be affected by a time-delay $h>0$. The connection between the plant and the PI+RI controller is defined by $u_p(t)=y_r(t)$ and $u_r(t)=-y_p(t-h)$.  The selection of the system state ${\mathbf x}(t)=({\mathbf x}_p(t),{\mathbf x}_r(t)) \in \mathds{R}^{n}$, with $n=n_p+2$, leads into the following description of the closed-loop system
\begin{equation}
\label{eq:RCS}
\left\{ \begin{array}{ll}
\dot{\mathbf{x}}(t) = A \x(t) + A_d \x(t-h),  & (t,\x(t)) \notin \mathcal{S}, \\
\x(t^+)=A_R \mathbf{x}(t), &  (t,\x(t)) \in \mathcal{S}, \\
\x(t)=\phi(t), & t\in [-h,0],
\end{array}	
\right.																		
\end{equation}
where  $\phi\in \mathcal{C}([-h,0],\mathds{R}^n)$ is the initial condition function, and the matrices $A$, $A_d$, and $A_R$ are given by
\begin{equation}\label{eq:ClosedLoopMatrices}
\begin{array}{cc c}
A=\left[ \begin{array}{c c}
A_p  & B_p C_r\\
0 & 0								
\end{array}	
\right], &A_d=\left[ \begin{array}{c c}
-k_p B_p C_p  & 0\\
-B_r C_p & 0								
\end{array}	
		\right], &A_R=\left[ \begin{array}{c c}
											I & 0\\
											0 & A_\rho								
											\end{array}	
		\right].
\end{array}	
\end{equation}
According to the resetting set of the reset controller PI+RI, the resetting set of the system (\ref{eq:RCS}) is defined by
\begin{equation}
\mathcal{S}=\{ (t,\x) \in \mathds{R} \times \mathds{R}^n: (t,\x_r)\in \mathcal{R}  \}.
\end{equation}

Due to the assumption on the resetting sequence, it is possible to conclude from the results in \cite{BL2000} that there exists a unique solution  ${\mathbf x}(t,\phi)$, or simply ${\mathbf x}(t)$, for $t \in [-h,\infty)$.

\subsection{Reset integrator as a sampled-data system}

In this section, we show how the reset integrator can be reformulated into the framework of sampled-data systems.
Consider a RI represented by the following equation
\begin{equation}
\label{eq:CI}
\left\{ \begin{array}{ll}
\dot{x}_{ri}(t)= u_{ri}(t),  & (t,x_{ri}(t),u_{ri}(t)) \notin \mathcal{R}_{ci} \\
x_{ri}(t^+)=0, &  (t,x_{ri}(t),u_{ri}(t)) \in \mathcal{R}_{ci}, \\
y_{ri}(t)=x_{ri}(t),
\end{array}	
\right.
\end{equation}
whose reset instants are defined by the resetting set $\mathcal{R}_{ci}$. Again, it is assumed that for every execution of the system there exists an strictly increasing sequence of nonzero instants defined by $\mathds{T}=\{t_k\}_{k\in\mathbb N}$ for which (\ref{def_T}) holds (note that a more general resetting set can be considered as long as the assumption holds). Then, from the input/output point of view, the above system is equivalent to the following sampled-data system
\begin{equation}
\label{eq:SampledCI}
\left\{ \begin{array}{ll}
\dot{x}_{s}(t)= u_s(t), \\
y_s(t)=x_{s}(t)-x_s(t_k), & t\in (t_k,t_{k+1}], \\
\end{array}	
\right.															
\end{equation}
where $t_k\in\mathds{T}$. The output of both systems are the same for any initial condition $x_{ri}(0)=x_s(0)$ and all input signal $e$. Therefore, a reset control system whose controller comprises a RI (\ref{eq:CI}) can be transformed into a sampled-data system by exchanging the RI with (\ref{eq:SampledCI}). In this way, the evolution of both systems are equivalent except for the states $x_i$ of (\ref{eq:CI}) and  $x_s$ of (\ref{eq:SampledCI}). However, it should be pointed out that $y_{ri}(t)=0$ implies $x_{ri}(t)=0$, but the counterpart implication for (\ref{eq:SampledCI}) is false. 

As a consequence, the internal stability of the reset control system does not imply the internal stability of the sampled-data system. However, we will use the reserve assumption, which is correct, to guarantee the asymptotic stability of the reset control system as a consequence of the asymptotic stability of the sampled-data system.

\subsection{PI+RI as a sampled-data system}
Let first us focus on the sampled-data formulation of the PI+RI controller. This can be done by simply considering the formulation (\ref{eq:SampledCI}) into the PI+RI, which results into
\begin{equation}
PI+RI:\left\{ \begin{array}{ccll}
\!\!\!\left[\begin{matrix}
\dot{x}_{i}(t)\\
\dot{x}_{s}(t)
\end{matrix}\right]&\!\!\!\!=&\!\!\!\! B_r u_r(t), \\
y_r(t)&\!\!\!\!=&\!\!\!\!(1\!-\!p_r)x_i(t)+p_r (x_s(t)\! -\! x_s(t_k))+ k_p u_r(t), &\!\! t\in (t_k,t_{k+1}].\\
\end{array}	
\right.
\end{equation}

A simple inspection of the above equation shows that the evolution of both $x_i$ and $x_s$ are equivalent, and thus, the realization of the system is non-minimal. However, a minimal realization of the PI+RI, equivalent (regarding input/output behavior) to (\ref{eq:PICI}), can be obtained by removing $x_i$ and carefully considering the initial condition $(x_{i}(0),x_{ri}(0))$ for (\ref{eq:PICI}). As a result, the PI+RI controller can be represented as follows:
\begin{equation}
\label{eq:PICIsampled}
PI+RI:\left\{ \begin{array}{ll}
\dot{x}_{s}(t)= k_i u_r(t), \\
y_r(t)=x_{s}(t)-p_r (x_i(0)-x_{ri}(0)) + k_p u_r(t), & t\in [0,t_1], \\
y_r(t)=x_{s}(t)-p_r x_s(t_k)+ k_p u_r(t), & t\in (t_k,t_{k+1}]. \\
\end{array}	
\right.																		
\end{equation}

Note that the equation $y_r(t)=x_{s}(t)\!-\!p_r (x_i(0)\!-\!x_{ri}(0)) + k_p u_r(t) $ guarantees the equivalence between the two realizations, (\ref{eq:PICI}) and (\ref{eq:PICIsampled}), of the PI+RI controller.

Finally, considering the realization (\ref{eq:PICIsampled}), the reset control system (\ref{eq:RCS}) is transformed into the following sampled-data system
\begin{equation} 
\label{eq:RCSsampled}
	\left\lbrace \begin{array}{lrll}
						\dot{X}(t) = \Lambda X(t)+ \Lambda_d X (t-h)+\Lambda {\bf u}(t),\\
						{\bf u}(t)=K X(t_k) \textrm{,} &  t \!\!\!&\in&\!\!\! (t_k,t_{k+1}],\\
						X(t)= \phi_X(t), & t\!\!\!&\in&\!\!\! [-h,0],
	\end{array}\right.													
\end{equation}
where $X(t)=({\bf x}_p(t),x_s(t))$, $\phi_X$ the initial condition and  the matrices $\Lambda$, $\Lambda_d$, and $K$  given by
\begin{equation}\label{eq:ClosedLoopAD_AdD_2}
\begin{array}{c c c}
\Lambda=\left[ \begin{array}{c c}
											A_p  & B_p \\
											0 & 0							
											\end{array}	
		\right], & \Lambda_d=\left[ \begin{array}{c c}
											-k_p B_p C_p  & 0\\
											-k_i C_p & 0								
											\end{array}	
		\right], & K=\left[ \begin{array}{c c}
											0 & 0\\
											0 & -p_r								
											\end{array}	
		\right].
\end{array}
\end{equation}

Note that, if $\phi_X(t)=[ \ I_{n_p+1} \ \ 0 \ ]\phi(t)$, $t\in [-h,0]$ and  $x_i(0)-x_{ri}(0)=[ \ 0_{n_p} \ \ 1 \ ] {\bf u}(0)$, then the evolution of the plant state, $\x_p$, is the same in both formulations and the states $x_i$ and $x_{ri}$ of (\ref{eq:PICI}) are directly retrieved from $X(t)$ since $x_i(t)=x_s(t)$ and $x_{ri}(t) =x_s(t)+[ \ 0_{n_p} \ \ 1 \ ] {\bf u}(t)$.

\section{Exponential stability analysis}
\label{sec:exponential_stability}

\subsection{Preliminaries}

This section provides an exponential stability criterion for the time-delay reset control system (\ref{eq:RCS}). First, it is proved the equivalence between the systems (\ref{eq:RCS}) and (\ref{eq:RCSsampled}), regarding the exponential stability.  Second, we propose sufficient conditions for the exponential stability of the sample-data system (\ref{eq:RCSsampled}), allowing to conclude on the stability of the system (\ref{eq:RCS}). As mentioned before, due to the particular structure of (\ref{eq:RCSsampled}), the results using the framework of sampled-data systems are not applicable.\\

The stability criterion developed in this section is based on a LK functional that depends on both the solution $X(t)$ and its derivative $\dot{X}(t)$. Therefore, it is assumed that the initial conditions are restricted to the space of absolutely continuous functions $\phi_X: [-h,0] \rightarrow \mathds{R}^n$ with square integrable derivative (see, e.g., \cite{F2014}), which is denoted by $W[-h,0]$ and its associated norm is
\begin{equation}
\label{eq:normaW}
\| \phi_X \|_W = \max_{\theta \in [-h,0]} \| \phi_X(\theta)\| + \left( \int^0_{-h} \| \dot{\phi}_X(\theta) \|^2 d\theta \right)^{\frac{1}{2}}.
\end{equation}

As discussed in \cite{F2014}, the stability results corresponding to continuous initial conditions are equivalent to the case of absolutely continuous initial conditions.

\begin{definition}
The trivial solution of system (\ref{eq:RCSsampled}) is said to be exponentially stable with decay rate $\alpha>0$, if for every $\epsilon>0$,  there exists $\delta=\delta(\epsilon)>0$ such that 
\begin{equation}
\| \phi_X \|_W < \delta \qquad \Rightarrow \qquad \| X(t) \| < \epsilon e^{-\alpha t}, \ \ \ \forall t\geq 0.
\end{equation}
\end{definition}

The exponential stability of the time-delay reset control system (\ref{eq:RCS}) is defined equivalently with $\| \phi \|_W < \delta $ and $\| \x(t) \| < \epsilon e^{-\alpha t}$, $t\geq 0$.

\begin{prop}
\label{prop:Prop1}
If the sampled-data system (\ref{eq:RCSsampled}) is exponentially stable then the reset control system (\ref{eq:RCS}) is exponentially stable. 
\end{prop}
\begin{pf}
The proof follows straightforwardly from the following relations:
\begin{equation}
\begin{array}{rclcc}
 \phi_X(t) &=& [ \ I_{n_p+1} \ \ 0 \ ] \phi(t), & t \in [-h,0],\\[0.8ex]
  x_i(0)-x_{ri}(0)&=&[ \ 0_{n_p} \ \ 1 \ ] {\bf u}(0), & \\[0.8ex]
  X(t)& =& [ \ I_{n_p+1} \ \ 0 \ ] {\bf x}(t), & t \geq 0,\\[0.8ex]
   x_{ri}(t)& = &x_s(t)-[ \ 0_{n_p} \ \ 1 \ ] {\bf u}(t), & t \geq 0.
   \end{array}
\end{equation} 
\qed

%
%
%
\end{pf}

The next result provides sufficient conditions for the exponential stability of the sampled-data system (\ref{eq:RCSsampled}), which are based on the existence of a LK functional and a family of functionals $\mathcal{V}_k$.  In these paper, the analysis is acheived thanks to the definition of a lifted function, $\chi_k \in  \mathcal{C}([0,T_k]\times [-h,0],\mathds{R}^{n})$, which is defined, for any reset instant $t_k \in \mathds{T}$ as follows:
\begin{equation}
\chi_k(\tau,\theta)= \chi(t_k+\tau+\theta).
\end{equation}
Using this notation, the dynamics of system  (\ref{eq:RCSsampled}) is rewritten as
\begin{equation} 
\label{eq:RCSsampled_chi}
\begin{array}{lrll}
						\dot{\chi}_k(\tau,0) = \Lambda \chi_k(\tau,0)+ \Lambda_d \chi_k(\tau,-h)+\Lambda K \chi_k(0,0), & &\forall \tau\in[0,T_k]&\forall k\in \mathbb N.\\

	\end{array}													
\end{equation}
%
%
%
%

The following proposition is an adaptation of the technique of looped functionals used in \cite{S2011,S2012}.

\begin{prop}
\label{prop:Prop2}
For given scalars $0<\alpha$ and $0\leq \mathcal{T}_m\leq \mathcal{T}_M$, consider a functional  $V:W[-h,0] \rightarrow \mathds{R}$ for which there exists some positive scalar $\mu_1$, $\mu_2$ such that, for all $\phi \in W[-h,0]$, the following relations hold
\begin{equation}\label{eq:GeneralExp_cond1}
\mu_1 \|\phi(0)\|^2  \leq V(\phi) \leq \mu_2 \| \phi \|^2_W.
\end{equation}

Moreover, if there exist a functional $\mathcal{V}_k: [0, T_k] \times \mathcal{C}([0,T_k]\times [-h,0],\mathds{R}^{n})  \rightarrow \mathds{R}$, for any $k$ in $\mathbb N$ verifying, first, the looping condition
\begin{equation}
\label{eq:GeneralExp_cond2}
e^{2\alpha T_k} \mathcal{V}_k(T_k, \chi_k)=\mathcal{V}_k(0,\chi_k),
\end{equation}
and 
\begin{equation} \label{eq:GeneralExp_cond3}
\mathcal{V}_k(0,\chi_k) \leq \eta V (\chi_k(0,\cdot)),
\end{equation}
\begin{equation} \label{eq:GeneralExp_cond4}
-\eta V(\chi_k(0,\cdot)) \leq e^{2\alpha \tau} \mathcal{V}_k(\tau,\chi_k)
\end{equation}
for some $\eta\geq0$ and $\tau \in [0,T_k]$, 
and, second,
\begin{equation}
\label{eq:GeneralExp_cond5}
\dot{W}(\sigma, \chi_k)=\frac{d}{d\sigma} \left( e^{2\alpha \sigma} ( V(\chi_k(\sigma,\cdot)) + \mathcal{V}_k(\sigma,\chi_k) ) \right) \leq 0
\end{equation}
for all $\sigma \in [0,T_k]$, then system (\ref{eq:RCSsampled}) is exponentially stable with decay rate $\alpha>0$ for any reset instants satisfying \eqref{def_T}.

\end{prop}
\begin{pf}

Consider the $k$-ith reset instant, then the integration of (\ref{eq:GeneralExp_cond5}) with respect to $\sigma$ over the interval $[0,\tau]$ lead into
\begin{equation}\label{eq:proof1_1}
e^{2\alpha \tau } \left( V(\chi_k(\tau,\cdot))  + \mathcal{V}_k(\tau,\chi_k) \right) - V(\chi_k(0,\cdot))    - \mathcal{V}_k(0,\chi_k)    \leq 0.
\end{equation}

Applying inequalities (\ref{eq:GeneralExp_cond2}), (\ref{eq:GeneralExp_cond3}), and (\ref{eq:GeneralExp_cond4}), it follows
\begin{equation}\label{eq:proof1_2}
V(\chi_{k}(\tau,\cdot)) \leq  (1+2\eta)e^{-2\alpha \tau} V(\chi_k(0,\cdot)),
\end{equation}
and
\begin{equation}\label{eq:proof1_3}
V(\chi_{k}(T_k,\cdot)) \leq  e^{-2\alpha T_k} V(\chi_k(0,\cdot)).
\end{equation}

Considering the previous reset instants and the Lipschitz continuity of the time-derivative of the state, there exists $K > (1+2\eta)$ such that
\begin{equation}\label{eq:proof1_4}
V(\chi_k(\tau,\cdot)) \leq K  e^{-2\alpha t} V(\phi_\chi ),
\end{equation}
where $t=t_k+\tau$. Inequality (\ref{eq:GeneralExp_cond1}) yield to 
\begin{equation}\label{eq:proof1_8}
\mu_1 \|  \chi_k(\tau,0) \|^2 \leq V(\chi_k(\tau,\cdot)) \leq K e^{-2\alpha t}  V(\phi_\chi )  \leq   K e^{-2\alpha t} \mu_2 \| \phi_\chi \|^2_W.
\end{equation}

Finally, for a given $\epsilon>0$, set $\delta < \sqrt{\frac{\mu_1}{K \mu_2}} \epsilon$, and thus, system (\ref{eq:RCSsampled}) is exponentially stable with decay rate $\alpha$. \qed

\end{pf}

\subsection{Main result}

The next result provides a sufficient criterion for the exponential stability of the sampled-data system (\ref{eq:RCSsampled}) expressed in terms of LMI. The following theorem is stated.

\begin{prop}
\label{prop:Prop3}
For given scalars $0<\alpha$ and $0\leq \mathcal{T}_m\leq \mathcal{T}_M$, the system (\ref{eq:RCSsampled}) is exponentially stable with decay rate $\alpha$ if there exist a matrix $P_N\in \mathbb{S}^{(N+1)n}$, matrices $S$, $R$, $Q_i$, $X_i$, $U_i \in \mathbb{S}^n_+$, matrices $Z_i \in \mathds{R}^n$ and $Y_i \in \mathds{R}^{(N+3)n \times n}$ such that the LMIs

\begin{equation} \label{eq:Prop4Cond1}
P_N + \frac{e^{-2 \alpha h} }{h}diag(0_n,S,3S,\ldots,(2N-1)S) > 0,
\end{equation}

\begin{equation}\label{eq:Prop4Cond2}
h_1(0) \Pi_{1,i} + h_2(0,\mathcal{T}_{i-1}) \Pi_{2,i}  + h_4(0,\mathcal{T}_{i}) N_2^\top X_i N_2  < 0,
 \end{equation}
 
 \begin{equation}\label{eq:Prop4Cond3}
h_1(0) \Pi_{1,i} + h_2(0,\mathcal{T}_{i}) \Pi_{2,i}  + h_4(0, \mathcal{T}_{i}) N_2^\top X_i N_2  < 0,
 \end{equation}

\begin{equation}\label{eq:Prop4Cond4}
\left[ \begin{array}{cc} h_1(\mathcal{T}_{i-1}) \Pi_{1,i} +  h_4(\mathcal{T}_{i-1},\mathcal{T}_{i}) N_2^\top X_i N_2 & h_3(\mathcal{T}_{i-1},\mathcal{T}_{i}) Y_i \\ \star & -h_3(\mathcal{T}_{i-1},\mathcal{T}_{i})  U_i
\end{array} \right] < 0,
\end{equation}

\begin{equation}\label{eq:Prop4Cond5}
\left[ \begin{array}{cc} h_1(\mathcal{T}_i) \Pi_{1,i} +  h_4(\mathcal{T}_i, \mathcal{T}_{i}) N_2^\top X_i N_2 & h_3(\mathcal{T}_i,\mathcal{T}_i) Y_i \\ \star & -h_3(\mathcal{T}_i,\mathcal{T}_i)  U_i
\end{array} \right] < 0
\end{equation}

\noindent hold for all $i\in \{1,2,\ldots, M\}$, where the functions $h_i$, $i=1,2,3,4$ are given by
\begin{equation}
\begin{array}{rcllrcl}
 h_1(\tau)&=& e^{2 \alpha \tau}, & & h_2(\tau,T)&=&\frac{e^{2 \alpha T} - e^{2 \alpha \tau} }{2 \alpha},\\
 h_3(\tau,T)&=&e ^{2 \alpha T} \left( \frac{e^{2 \alpha \tau} - 1 }{2 \alpha} \right),&& 
 h_4(\tau,T)&=&\textcolor{black}{\left( \left(e^2-4\right) e^{\alpha T}+ 1 \right) e^{2\alpha \tau} +2e^{2\alpha T}},
\end{array}
\end{equation}
 and where the matrices $\Pi_{j,i}$, $i=1,\dots m$, $j=1,2$ are given by
 \begin{equation}
 \begin{array}{lcl}
 \begin{split}
 \Pi_{1,i}&=He\left( G^\top P H \right) + \Sigma_N + h^2 F^\top R F- \Gamma_{N}^\top R_N \Gamma_{N} \\&-N_{12}^\top Q_i N_{12} -  He\left(  N_2^\top Z_i N_2 \right) + He\left(  Y_i N_{12} \right) \end{split}
 \\
 \\ 
 \Pi_{2,i} =F^\top U_i F + He\left( F^\top Q_i N_{12} \right)  +  He\left( F^\top Z_i N_2\right)  \\[0.3cm]
\end{array}
\end{equation}
with $N_1\!=\!\left[ I_n \ 0_{n,n(N\!+\!1)} \right]$, $N_2\!=\!\left[0_{n,n(N\!+\!1)}  \ I_n \right]$, $N_{12}\!=\!N_1\!-\!N_2$ and where
\begin{equation}
\label{eq:functions_and_matrices}
\begin{array}{rll}

 R_N=&e^{-2\alpha h}\ diag( R, \ 3R,\  \ldots\ ,\  &\!\!\!\!\!\!(2N\!+\!1)R), \\[0.3cm] 

\Sigma_N=&\!\!\!\! diag ( S,-e^{-2\alpha h}S,0_{(N+1)n} ) ,

&G =\left[ \begin{array}{cccc} I_n & 0_n & 0_{n,nN}  &0_n \\ 0_{nN,n} & 0_{nN,n} & h I_{nN} & 0_n  \end{array} \right], \\[0.5cm]

F =&\!\!\!\! \left[ \Lambda \ \Lambda_d \ 0_{n,nN} \ \Lambda K \right],&
 \Gamma(k) =\left[ I \ \ (-I)^{k\!+\! 1} \  \gamma^0_{k}I \  \ldots \ \gamma_{k}^{N-1} I  \ 0_n  \right] , \\[0.3cm]

H =& \!\!\!\!\left[ \! \!\begin{array}{c} F\\ [0.3cm]
\Gamma (0)  \\ 
\vdots \\
\Gamma (\!N\!\!-\!\!1\!) \end{array}\!\!\right]\!\!,\  \Gamma_N\left[ \! \!\begin{array}{c} 
 \Gamma(0) \\  \Gamma(1) \\ \vdots \\ \Gamma(N) \end{array} \right] ,
 &\gamma_{k}^i = \left\lbrace \begin{array}{c c r} -(2i+1)(1- (-1)^{k+i}), & i\leq k ,\\
0,&  i>k.
\end{array}\right.
 \end{array}
\end{equation}

\end{prop}
\begin{pf} 
The proof of this proposition is based on the application of Proposition \ref{prop:Prop2}. Therefore, for the sake of clarity, the proof is  divided into three steps to show how each condition of \ref{prop:Prop2} can be verified. These steps are described as follows:
\begin{description}
\item[1)] Construction of the functional $V$ and an LMI condition ensuring the satisfaction of inequality \eqref{eq:GeneralExp_cond1}; 
\item[2)] Construction of the looped-functional $\mathcal V_k$, satisfying the looping condition \eqref{eq:GeneralExp_cond2} and  conditions \eqref{eq:proof1_2} and \eqref{eq:proof1_3};  
\item[3)] Design of LMI conditions guaranteeing \eqref{eq:GeneralExp_cond5}.
\end{description}

\textbf{1) Contruction of $V$:}
In the literature, several class of functionals have been studied, leading to a wide class of stability conditions. In this paper, we focus on a recent development on the stability analysis of time-delay systems based on Bessel inequality and Legendre's polynomials provided in \cite{SG2014}. Therein, the following LK functionals has been introduced
\begin{equation}\label{eq:LK}
 \begin{split}
V(\chi_k(\tau,\cdot))&=\tilde{ \chi}^\top_{k} (\tau)P_N\tilde{ \chi}_{k}(\tau) +\int^0_{-h} e^{2 \alpha s} \chi_k^\top(\tau,s)S\chi_k(\tau, s) ds \\ &
+h \int^0_{-h} \int^0_{\beta}  e^{2 \alpha s}  \dot{ \chi}_k^\top (\tau,s) R  \dot{ \chi}_k (\tau,s) ds d\beta,
\end{split}
\end{equation}
where the augmented state $\tilde{\chi}_{k}(\tau)$ is given by
\begin{equation}
\tilde{\chi}_{k}(\tau)= \left[ \begin{array}{c}  \chi_k(\tau,0) \\ \int^0_{-h} L_0(s) \chi_k(\tau,s) ds  \\ \vdots \\ \int^0_{-h} L_{N-1}(s) \chi_k(\tau, s) ds \end{array} \right],
\end{equation}
where we use the Legendre polynomials considered over the interval $[-h,0]$ and given by
\begin{equation}
\label{eq:LegendrePolynomial}
L_k= (-1)^k \sum^{k}_{l=0} (-1)^l \left(\begin{smallmatrix}k\\l\end{smallmatrix}\right)\left(\begin{smallmatrix}k+l\\l\end{smallmatrix}\right)\left(  \frac{u+h}{h} \right)^l,\qquad \forall k \in \mathds{N},
\end{equation}
where the notation $\left(\begin{smallmatrix}a\\b\end{smallmatrix}\right)$ with $a,b$ in $\mathbb N$ denotes the binomial coefficients. The potential of using of these polynomials has been demonstrated in \cite{SG2014} and arisen from their orthogonal properties, the simple evaluation of the boundary values and the recursive relation to compute their derivative using the same polynomial basis of lower degree. All these details are already provided in \cite{SG2014} and are therefore omitted in this paper. 

In order to prove that condition \eqref{eq:GeneralExp_cond1} holds for this Lyapunov functionals is inspired from  \cite{SG2014}. Firstly, considering the bounds $e^{-2\alpha h} \leq e^{2\alpha s} \leq 1$ for $s\in[-h,0]$, then, since $R>0$, we have
\begin{equation}\label{eq:LK1}
 \begin{array}{lcl}
V(\chi_k(\tau,\cdot))&\leq&\tilde{ \chi}^\top_{k} (\tau)P_N\tilde{ \chi}_{k}(\tau) +e^{2 \alpha h}\int^0_{-h}  \chi_k^\top(\tau,s)S\chi_k(\tau, s) ds \\ 
&\leq & \tilde{ \chi}^\top_{k} (\tau)\left(P_N+ \frac{e^{-2 \alpha h} }{h}diag(0_n,S,3S,\ldots,(2N-1)S)\right) \tilde{ \chi}_{k}(\tau).
\end{array}
\end{equation}
The last inequality has been derived accordingly to \cite{SG2014} with the use of the Bessel-Legendre inequality.
Then, if the condition \eqref{eq:Prop4Cond1} holds, it directly follows that the LK functional (\ref{eq:LK}) satisfies condition (\ref{eq:GeneralExp_cond1}).

\bigskip
\textbf{2) Contruction of $\mathcal V_k$:}
Consider now the following functionals
\begin{equation}
\begin{split}
\mathcal{V}_k(\tau, \chi_k)&= f_1(\tau,T_k) \left( \zeta_k^\top (\tau)   Q(T_k) \zeta_k(\tau)+He( \zeta_k^\top (\tau)   Z(T_k) \chi_k(0,0) ) \right) \\
&+ f_1(\tau,T_k) \int^\tau_0 \dot{\chi}_k^\top(s,0) U(T_k) \dot{\chi}_k(s,0) ds \\
&+f_2(\tau,T_k)  {\chi}_k^\top(0,0) X(T_k) {\chi}_k (0,0) + \gamma e^{-2\alpha \tau} \| \chi_k(0,0)\|^2
\end{split}
\end{equation}
for some $\gamma>0$, and where $\zeta_k(\tau)={\chi}_k(\tau,0)-{\chi}_k(0,0)$, $i \in \{ 1,\ldots, M \}$ satisfies $T_k \in [\mathcal{T}_{i-1}, \mathcal{T}_i ]$ and 
\begin{equation}
f_1(\tau,T_k)=\frac{e^{2 \alpha (T_k-\tau) }-1} {2\alpha},
\end{equation} 
\begin{equation}
\textcolor{black}{f_2(\tau,T_k)=\frac{e^{-2 \alpha \tau}}{2\alpha}\left( \left( e^{2 \alpha T_k }-1\right)\left( e^{2\alpha \tau }-1\right)  - \left( e^{2 \alpha \tau }-1\right) ^2
\right)},
\end{equation}
and where 
\begin{equation}
\begin{array}{lcll}
Q(T_k)&=&Q_i,\qquad &\mbox{ if } T_k\in[\mathcal T_{i-1}, \ \mathcal T_i],\\
Z(T_k)&=&Z_i,\qquad &\mbox{ if } T_k\in[\mathcal T_{i-1}, \ \mathcal T_i],\\
U(T_k)&=&U_i,\qquad &\mbox{ if } T_k\in[\mathcal T_{i-1}, \ \mathcal T_i],\\
X(T_k)&=&X_i,\qquad &\mbox{ if } T_k\in[\mathcal T_{i-1}, \ \mathcal T_i].
\end{array}
\end{equation}
This selection of functionals represents a new discretization method with respect to the literature. The parameters of the looped-functional are piecewise constants and have jumps only at the sampling instants. However, the due to the looping conditions, the effects of these jumps vanishes and the functional $\mathcal V_k$ is still continuous.

The functionals $\mathcal{V}_k$ satisfy condition  (\ref{eq:GeneralExp_cond2}) for all $k>0$ since $e^{2\alpha T_k}\mathcal{V}_k(T_k, \chi_k)=\mathcal{V}_k(0, \chi_k)= \gamma \| \chi_k(0,0) \|^2$. Now consider the following bound of the LK functional
\begin{equation}
\label{eq:LKbound}
\mu_1 \| \chi_k(\tau,0) \|^2  \leq V(\chi_k(\tau,\cdot)),
\end{equation}
then by taking $\tau=0$, it can be obtained the following inequality
\begin{equation}
\mathcal{V}_k(0,\chi_k) \leq \frac{\gamma}{\mu_1} V(\chi_k(0,\cdot)).
\end{equation}
Therefore, the functionals $\mathcal{V}_k$ and $V$ satisfy condition  (\ref{eq:GeneralExp_cond3}) for  $\eta \geq  \frac{\gamma}{\mu_1}$. 

Consider now the following term of the functionals $\mathcal{V}_k$
 \begin{equation}
  \label{eq:proof_positive_term}
 f_1(\tau,T_k) \left( \zeta_k^\top (\tau)   Q_i \zeta_k(\tau)+He( \zeta_k^\top (\tau)   Z_i \chi_k(0,0) ) \right)  +\gamma e^{-2\alpha \tau} \| {\chi}_k (0,0)\|^2.
\end{equation}
 The above term is nonnegative if the following LMI holds for $\tau \in [0,T_k]$, all $k>0$ and a sufficiently large $\gamma$
 \begin{equation}
 \label{eq:proof_LMI}
 \textcolor{black}{\left[ \begin{array}{ll}
  f_1(\tau,T_k) Q_i &  f_1(\tau,T_k) Z_i \\ \star & \gamma e^{-2 \alpha \tau}  I_n
  \end{array} \right] \geq 0.}
 \end{equation}
Note that $Q_i$ is positive definite, then applying the Schur complement, the above LMI holds if 
 \begin{equation}
 \gamma I_n - e^{2\alpha\tau}  f_1(\tau,T_k) Z_i^\top Q_i^{-1} Z_i  \geq 0.
 \end{equation}
 Since $e^{2\alpha\tau}  f_1(\tau,T_k)$ is bounded for $\tau \in [0, T_k]$ and $ T_k \in [\mathcal{T}_m, \mathcal{T}_M]$, there exists a sufficiently large $\gamma$ such that the LMI (\ref{eq:proof_LMI}) holds for all $i \in \{1,\ldots, M\}$. In addition, the non-negativeness of the term (\ref{eq:proof_positive_term}) and $U_i>0$ imply that  $\gamma$ might be chosen sufficiently large such that 
 \begin{equation}
 -\gamma \| {\chi}_k^\top(0,0) \|^2 \leq e^{2\alpha \tau }f_2(\tau,T_k) {\chi}_k^\top(0,0) X_i {\chi}_k (0,0)  \leq  e^{2\alpha \tau } \mathcal{V}_k (\tau,\chi_k).
 \end{equation}
Hence, considering equation (\ref{eq:LKbound}), it follows that condition  (\ref{eq:GeneralExp_cond4}) is satisfied for $\eta \geq \frac{\gamma}{\mu_1}$.

\textcolor{black}{Note that $\gamma$ does not play any role in the conditions of Proposition \ref{prop:Prop3}, and thus, it is not necessary to set a value for $\gamma$. The previous reasoning proves that there exists a suitable value for $\gamma$, which is enough for the functionals $\mathcal{V}_k$ to fulfill the conditions in Proposition  \ref{prop:Prop2}.
}

\bigskip
 
\textbf{3) Design of LMI conditions guaranteeing \eqref{eq:GeneralExp_cond5}:} Following \cite{SG2014}, the computation of the time-derivative of $e^{2 \alpha \tau } V(\chi_k(\tau,\cdot))$ along the trajectories of the system (\ref{eq:RCSsampled}) is given by
\begin{equation}
\begin{split}
\frac{d}{d\tau} \left( e^{2\alpha \tau} V(\chi_k(\tau,\cdot)) \right) &=
 e^{2 \alpha \tau}   \Big( He(\tilde{ \chi}^\top_{k} (\tau )P_N \dot{ \tilde{\chi}}_{k}(\tau)) + 2 \alpha \tilde{ \chi}^\top_{k} (\tau )P_N  \tilde{\chi}_{k}(\tau)  \\ &+   \chi_k^\top(\tau,0) S \chi_k(\tau,0)  -  e^{- 2 \alpha h} \chi_k^\top(\tau,-h) S \chi_k(\tau, -h)   \\
&+ h^2 \dot{ \chi}_k^\top(\tau,0) R \dot{ \chi}_k(\tau, 0) 
-h \int^0_{-h} e^{2\alpha s}\dot{ \chi}_k^\top(\tau, s)R \dot{ \chi}_k(\tau, s) ds  \Big).
\end{split}
\end{equation}

The first step consists in computing the time-derivative of the augmented state $\tilde{\chi}_k$. Following \cite{SG2014} and the differentiation properties of the Legendre polynomials, we have 
$$\dot{\tilde{\chi}}_k= H  \xi_k(\tau),$$
where the matrix $H$ is given in \eqref{eq:functions_and_matrices} and where the vector $\xi_k$ is another augmented vector, which gathers all the relevant information for the analysis and which is given by
\begin{equation}
\xi_{k}(\tau)= \left[ \begin{array}{c}  \chi_k(\tau,0) \\ \chi_k(\tau,-h) \\  \frac{1}{h} \int^0_{-h} L_0(s) \chi_k(\tau,s) ds  \\ \vdots \\ \frac{1}{h} \int^0_{-h} L_{N-1}(s) \chi_k(\tau,s) ds \\  \chi_k(0,0) \end{array}   \right].
\end{equation}
 
Moreover noting that  $\tilde{ \chi}_k(\tau) = G \xi_k(\tau)$, the following expression of the time-derivative of  $e^{2\alpha \tau} V(\chi_k(\tau,\cdot))$ is rewritten as follows:
\begin{equation}
\frac{d}{d\tau} \left( e^{2\alpha \tau} V(\chi_k(\tau,\cdot)) \right) =
 e^{2 \alpha \tau}   \left( \xi^\top_k(\tau) \Phi   \xi_k(\tau)   - h \mathcal{I} (\dot{\chi}_k(\tau,\cdot)) \right)
\end{equation}
with 
\begin{equation}
\mathcal{I} (\dot{\chi}_k(\tau,\cdot)) = \int^0_{-h} e^{2\alpha s}\dot{ \chi}_k^\top(\tau, s)R \dot{ \chi}_k(\tau, s) ds.
\end{equation}
 
The next step consists in expressing the time-derivative of $\mathcal{W}=e^{2\alpha \tau } ( V +  \mathcal{V}_k)$. The calculations lead to
\begin{equation}
\begin{split}
\dot{\mathcal{W}}(\tau, \chi_k)&= h_1(\tau) \left( \xi_{k}^\top (\tau) \Phi \xi_{k} (\tau) -h \mathcal{I}(\dot{ \chi }_k(\tau,\cdot)) \right)\\
& - h_1(\tau) \big(   \zeta_k^\top (\tau)   Q_i \zeta_k(\tau)+He( \zeta_k^\top (\tau)   Z_i \chi_k(0,0) )   \big) \\
&- h_1(\tau) \int^\tau_0 \dot{\chi}_k^\top(s,0) U_i \dot{\chi}_k(s,0) ds \\
&+h_2(\tau,T_k) \left( He(\dot{\chi}^\top_k (\tau,0) Q_i \zeta_k(\tau)) + He(\dot{\chi}^\top_k (\tau,0) Z_i \chi_k(0,0)) \right)\\
&+h_2(\tau,T_k) \left( \dot{\chi}^\top_k (\tau,0) U_i \dot{\chi}_k (\tau,0) +  \dot{\chi}^\top_k (0,0) U_i \dot{\chi}_k (0,0) \right)\\
&+\left( ( e^{2 \alpha T_k}+1 ) e^{2\alpha \tau} -2e^{4\alpha \tau } \right) \chi^\top_k (0,0) X_i \chi_k(0,0).
\end{split}
\end{equation}

In the previous expression, there are two integral quadratic terms that cannot be included directly into an LMI. Therefore, the integral terms in the right-hand side of the above equation are bounded by the following two inequalities
\begin{equation}\label{eq:IntegralInequalityFred}
\mathcal{I}(\dot{ \chi }_k(\tau,\cdot)) \geq \frac{1}{h} \xi_k(\tau)^\top \left[ \sum_{j=0}^N (2j+1) \Gamma_N^\top(j) R \Gamma_N(j)   \right] \xi_k(\tau),
\end{equation}
\begin{equation}\label{eq:IntegralInequality}
-\int^\tau_0 \dot{\chi}_k^\top(s,0) U \dot{\chi}_k(s,0) ds \leq 2 \xi_{k}^\top (\tau) Y  \zeta_k(\tau) \\ 
+\tau \xi_{k}^\top (\tau)  Y U^{-1} Y^\top \xi_{k} (\tau).
\end{equation}

The first inequality was proposed in \cite{SG2014} based on the properties of Legendre polynomials (\ref{eq:LegendrePolynomial}) and Bessel's inequality. The second inequality can be found for instance in \cite{S2011,S2012}. In addition, we can retrieve the following inequalities
\begin{equation}
\label{eq:h1_h3_bound}
\begin{array}{lr}
\tau h_1(\tau) \leq h_3(\tau,T),  & \tau \in [0,T],
\end{array}
\end{equation}
\begin{equation}
\label{eq:h4_bound}
\begin{array}{lr}
\textcolor{black}{
( e^{2 \alpha T_k}+1 ) e^{2\alpha \tau} -2e^{4\alpha \tau }  \leq  h_4(\tau,\mathcal{T}_i), } & \left\{ \begin{array}{l}  T_k \in [\mathcal{T}_{i-1},\mathcal{T}_i], \\ \tau \in [0,T_k]. \end{array} \right.
\end{array}
\end{equation}

Using (\ref{eq:IntegralInequalityFred}), (\ref{eq:IntegralInequality}), (\ref{eq:h1_h3_bound}), and (\ref{eq:h4_bound}), the time-derivative of the functional $\mathcal{W}=e^{2\alpha \tau } ( V +  \mathcal{V}_k)$ is bounded by $\dot{\mathcal{W}}(\tau, \chi_k)  \leq \xi_{k}^\top (\tau) \Psi_i(\tau,T_k) \xi_{k} (\tau) $ with
\begin{equation}
\Psi_i(\tau,T_k) = h_1(\tau) \Pi_{1,i} + h_2(\tau,T_k) \Pi_{2,i} +  h_3(\tau,\mathcal{T}_i) Y_i U_i^{-1} Y_i^\top + h_4(\tau,\sigma, \mathcal{T}_i) N_2^\top X_i N_2 
\end{equation}
for some $i\in\{1,\ldots,M\}$. Note that $\Psi_i(\tau,T_k)$ is linear with respect to the term $e^{2\alpha \tau}$, and thus it is necessary and sufficient to ensure the negativity at the edges. This leads to $\Psi_i(0,T_k)<0$ and $\Psi_i(T_k,T_k)<0$. In addition, $\Psi_i(\tau,T_k)$ is also linear with respect to $e^{2\alpha T_k}$, hence applying the same convexity arguments on $T_k$, it is obtained 
\begin{equation}
\begin{array}{l}
\Psi_i(0,\mathcal{T}_{i-1})<0, \\[0.3em]
\Psi_i(0,\mathcal{T}_i)<0,  \\[0.3em]
\Psi_i(\mathcal{T}_{i-1},\mathcal{T}_{i-1})<0, \\[0.3em]
\Psi_i(\mathcal{T}_{i},\mathcal{T}_i)<0,
\end{array}
\end{equation}
which are rewritten (using the Schur complement) as (\ref{eq:Prop4Cond2}), (\ref{eq:Prop4Cond3}), (\ref{eq:Prop4Cond4}), and (\ref{eq:Prop4Cond5}), respectively, and the proof is complete. \qed
\end{pf}

\section{Examples}
\label{sec:Examples}

\subsection{Destabilization of a stable base system}
\label{subsec:example1}

\begin{figure}
\begin{center}
  \includegraphics[scale=1]{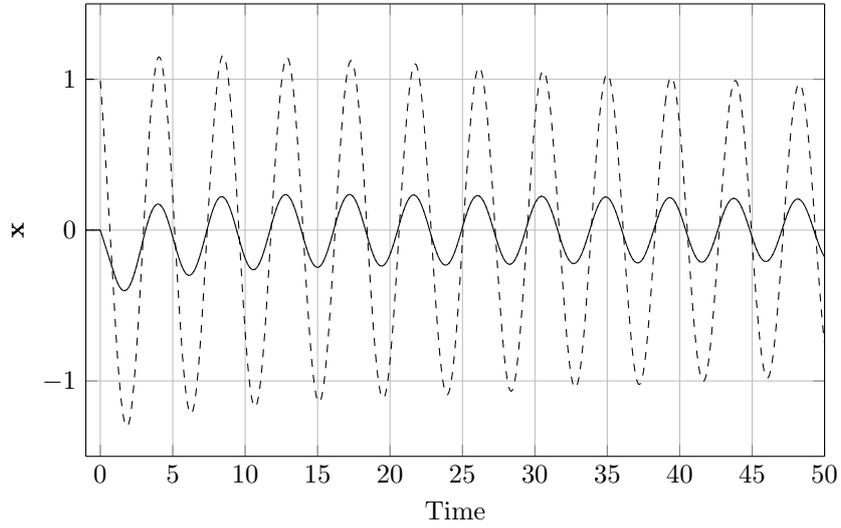}
  \caption{Trajectory of the base system of the Example \ref{subsec:example1}.}
  \label{fig:Example1_base}
\end{center}
\end{figure}

\begin{figure}
\begin{center}
  \includegraphics[scale=1]{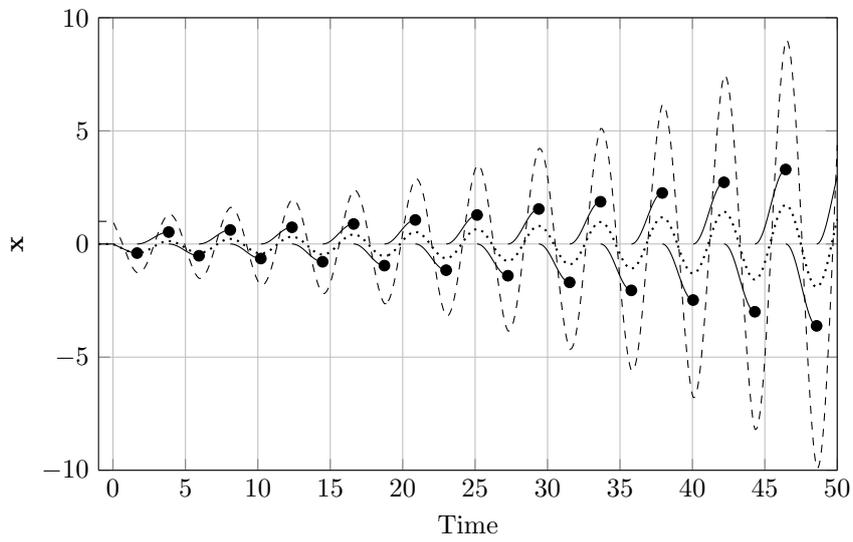}
  \caption{Trajectory of the reset control system of the Example \ref{subsec:example1}.}
  \label{fig:Example1_reset}
\end{center}
\end{figure}

A multitude of works have shown the advantages of the reset control to stabilize unstable systems (only non delayed systems). Consequently, one may wrongly think that the reset control system must be stable at least whenever its base system is stable. Several examples in \cite{r_HZCH1997, r_BCB2011} have been proved (for non delayed systems) that intuition fails, and reset actions can destabilize a stable system. This example shows that reset actions can destabilize a stable base system, and how the addition of the bounds $\mathcal{T}_m$ and  $\mathcal{T}_M$ lead into an exponentially stable reset control system. Consider a reset control system  that consists of a pure integrator plant, $P(s)=1/s$, controlled by the PI+RI (\ref{eq:PICI}) with $k_p=1.4$, $k_i=0.3$, $p_r=0.5$ and a zero crossing resetting law (a PI+CI controller). The closed-loop matrices defined in (\ref{eq:ClosedLoopMatrices}) are given by 
\begin{equation}
\begin{array}{ccc}
A=\left[ \begin{array}{ccc} 0 & 0.5 & 0.5 \\ 0 & 0 &0 \\  0 & 0 &0
\end{array}\right],  & 
A_d=\left[ \begin{array}{ccc} -1.4 & 0 & 0\\ -0.3 & 0 & 0\\ -0.3 & 0 & 0
\end{array}\right], & 
A_R=\left[ \begin{array}{ccc} 1 & 0 & 0 \\ 0 & 1 & 0 \\ 0 & 0 & 0
\end{array}\right].
\end{array}
\end{equation}
The base system with a minimal realization has been proved to be asymptotically stable for a time-delay $h=1$ by using the results in \cite{GP2006}. Fig. \ref{fig:Example1_base} shows how the trajectory of the base system with initial condition $\phi(t)=(1,0)$, $t\in[-1,0]$ converges to the zero solution. However, Fig. \ref{fig:Example1_reset} shows the solution of the reset control system for the same initial condition, and it can be seen that the trajectory is divergent. Now consider that the reset instants are forced to satisfy $\mathcal{T}_m\leq t_{k+1}-t_k \leq 1= \mathcal{T}_M$, then Table \ref{tab:Example1_table} shows the minimum $\mathcal{T}_m$ for several values of $M$, such that Propositions \ref{prop:Prop1} and \ref{prop:Prop3}  guarantee the exponential stability of the reset control system with decay rate $\alpha=1e^{-6}$. It is important to note that values of $N$ greater than two do not provide further improvements for this particular example.

\begin{table}
\begin{center}
\begin{tabular} {l c c c c c }
\toprule
$M$ & 1 & 3 & 5 & 10 & 50 \\
\midrule
\\[-2ex]
$\mathcal{T}_m$& 0.94 & 0.80 & 0.67 & 0.5  & 0.41\\[1ex]
\bottomrule
\end{tabular}
\end{center}
\caption{Effect of the parameter $M$ on the lower bound $\mathcal{T}_m$ for the Example \ref{subsec:example1}. ($\mathcal{T}_M=1$, $N=2$, and $\alpha=10^{-6}$).}
\label{tab:Example1_table}
\end{table}

\subsection{Stabilization of an unstable base system}
\label{subsec:example2}
\begin{figure}
\begin{center}
  \includegraphics[scale=1]{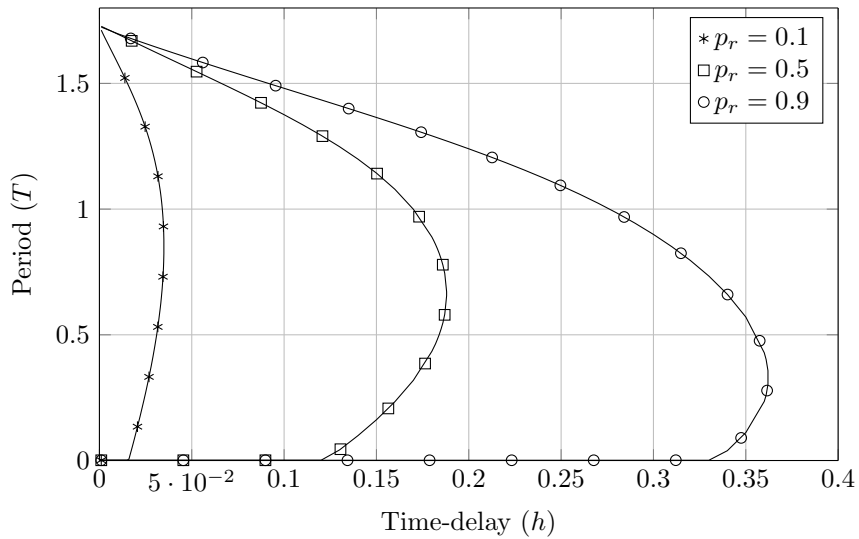}
  \caption{Allowable reset period ($T$) vs. time-delay ($h$) for several values of the reset ratio ($p_r$) for the Example \ref{subsec:example2}. ($N=2$, $M=1$, and $\alpha=10^{-6}$).}
  \label{fig:Example2_period_vs_delay}
\end{center}
\end{figure}

\begin{figure}
\begin{center}
  \includegraphics[scale=1]{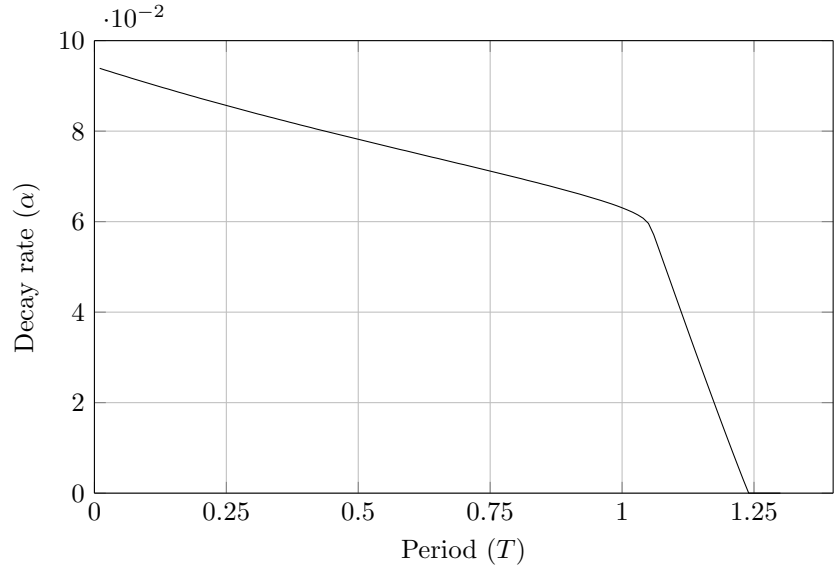}
  \caption{Bound of the decay rate ($\alpha$) vs. reset period ($T$) for the Example \ref{subsec:example2}. ($N=2$ and $M=1$).}
  \label{fig:Example2_decay_vs_period}
\end{center}
\end{figure}

\begin{figure}
\begin{center}
  \includegraphics[scale=1]{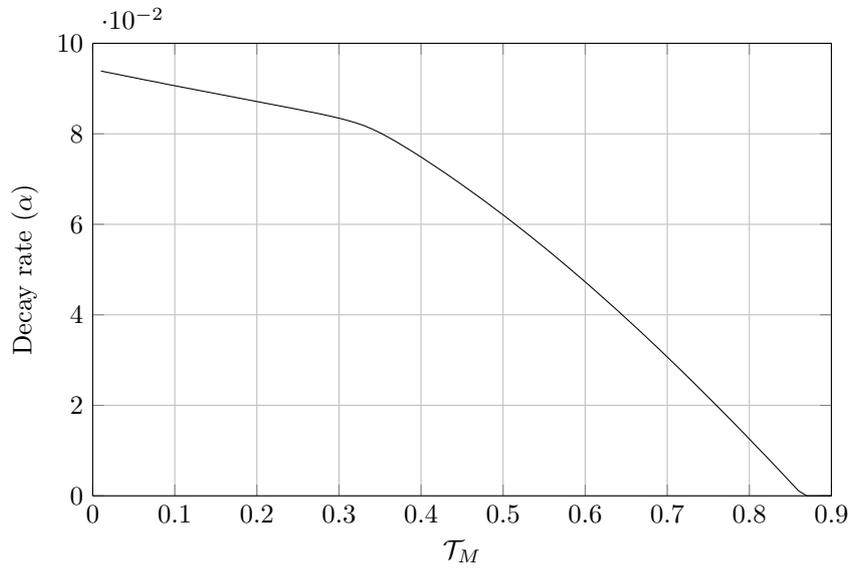}
  \caption{Bound of the decay rate ($\alpha$) vs. asynchronous reset period ($\mathcal{T}_m=0.01\mathcal{T}_M$) for the Example \ref{subsec:example2}. ($N=2$ and $M=1$).}
  \label{fig:Example2_decay_vs_period_asyn}
\end{center}
\end{figure}

Consider a reset control system composed by a plant P with matrices
\begin{equation}\label{eq:Ex1_matrices}
\begin{array}{ccc}
 A_p=\left[ \begin{array}{c c} 0 & 0\\ 1 & 0.5 \end{array} \right] ,& B_p= \left[ \begin{array}{c} 1\\ 1 \end{array} \right], & C_p = \left[ \begin{array}{c c} 0 & 1 \end{array} \right],	
\end{array}
\end{equation}
and a PI+RI compensator with $k_p=1$ and $k_i=1$. The base closed-loop system is not stable independently of the time-delay. However, the exponential stability of the reset control system can be guaranteed by the proposed results with a proper bounds $\mathcal{T}_m$ and $\mathcal{T}_M$.  First, we consider that the reset actions are periodic with period $\mathcal{T}_m=T_k=\mathcal{T}_M=T$, for all $k>0$. In Fig. \ref{fig:Example2_period_vs_delay}, it can be seen the effect of the reset ratio and the time-delay on the reset period. The results show that the time-delay can increase by increasing the reset ratio, while the exponential stability is guaranteed, In addition, Fig. \ref{fig:Example2_decay_vs_period} shows how the reset period affects the decay rate of the system. Finally, we consider asynchronous reset instants with bounds $\mathcal{T}_m=0.01 \mathcal{T}_M (>0)$. The effect  of $\mathcal{T}_M$ on the decay rate is shown in Fig. \ref{fig:Example2_decay_vs_period_asyn}.

\section{Conclusions}
\label{sec:conclusions}

New theoretical conditions through LMI conditions have been proved to ensure the exponential stability of a class of time-delay reset control systems. The time-delay reset control system is constituted by a PI+RI controller (generalization of the PI+CI controller) and a time-delay plant. The proposed approach is based on the extension to results issues from the sampled-data system framework. This work lets some questions open. In particular it could be interesting to propose conditions in order to design the controller, by adding for example some performance condition (such as rejection of perturbations).

\section*{Acknowledgements}
This work has been supported by Ministerio de Econom{\'i}a e Innovaci{\'o}n of Spain under project DPI2013-47100-C2-1-P (including FEDER co-funding) and the ANR project LimICoS contract number 12-BS03-005-01.

\bibliography{references}

\end{document}